\documentstyle[prd,aps,preprint]{revtex}
\begin{document}
\draft

%
%

\preprint{Nisho-01/1} \title{Skyrmion $\leftrightarrow$ pseudoSkyrmion Transition\\
in Bilayer Quantum Hall States at $\nu =1$}
\author{Aiichi Iwazaki}
\address{Department of Physics, Nishogakusha University, Shonan Ohi Chiba
  277,\ Japan.} \date{August 1, 2001} \maketitle
\begin{abstract}
Bilayer quantum Hall states at $\nu =1$ have been demonstrated to
possess a distinguished state with interlayer phase coherence.
The state has both excitations of Skyrmion with spin and 
pseudoSkyrmion with pseudospin.
We show 
that Skyrmion $\leftrightarrow$ pseudoSkyrmion transition arises 
in the state by changing
imbalance between electron densities in both layers;
PseudoSkyrmion is realized at balance point, while Skyrmion is
realized at large imbalance.
The transition can be seen by observing the dependence of activation
energies on magnetic field parallel to the layers.

\end{abstract}
\hspace*{0.3cm}
\pacs{PACS 73.43.-f,73.21.-b,71.10.Pm,12.39.De \\ 
Quantum\,Hall\,Effects,\,\, Skyrmion,\,\,
Interlayer\,Phase\,Coherence
\hspace*{1cm}}
It has recently been paid great interests to bilayer
quantum Hall states\cite{rev} at the filling factor $\nu=1$. Especially,
interlayer phase coherence and Goldstone boson
have been observed in recent experiments \cite{cohe,gold}.
Their existence 
has been originally derived \cite{iwa,zee,macdonald}
from the similarity between superconducting states and quantum
Hall states; both of them are characterized as 
the condensed states of electron pairs or composite electrons\cite{iwa1}.  
Furthermore, Skyrmion excitations \cite{skyrmion} 
have been shown \cite{sawada} to exist
in bilayer quantum Hall states at $\nu=2$. 
Before these observations, 
several measurements 
on the states have been performed\cite{murphy,sawada1,sawada2} by imposing
parallel magnetic field or by changing imbalance 
between electron densities in both layers. As a result it have been shown that 
the states have various types of excitations.

In this paper we discuss excitations of Skyrmion and
pseudoSkyrmion\cite{ezawa} in the bilayer quantum Hall state 
with the interlayer phase coherence at $\nu=1$.  
Particularly, we show that pseudoSkyrmion
is realized in the state with equal electron densities in two layers and
that a transition from the 
pseudoSkyrmion to Skyrmion arises by changing imbalance of the electron densities.

Skyrmion excitation has real spin components, while pseudoSkyrmion
excitation has pseudospin components ( pseudospin up  
and down components are 
given by the number of  electrons in front-layer
and in back-layer, respectively ).
Suppose that the density, $\rho_f$, in the front-layer is equal to the 
density, $\rho_b$, in the back-layer ( balanced state ).
With an appropriate interlayer distance $d\simeq l$ ( 
$l$ being magnetic length ), 
a state with interlayer phase coherence
is realized at $\nu=1$. We assume that
tunneling energy, $\Delta_{sas}\sim O$(K)  
between the two layers and Zeeman energy, $g\mu B\sim O$(K),
are much smaller than a typical Coulomb energy, 
$e^2/\epsilon l\sim 100$K,
where $g$ ( $\mu $ ) is electron $g$ factor ( Bohr magneton )
and $\epsilon\simeq 13 $ is the dielectric constant. Then,
we can show that the pseudoSkyrmion is an excitation 
with the lowest energy in the balanced state. 
As making the imbalance, $\sigma=(\rho_f-\rho_b)/(\rho_f+\rho_b)$,  
larger, the energy of the pseudoSkyrmion increases, 
while the energy of the Skyrmion 
does not change so much. Thus, at a critical point $\sigma_c$ of the 
imbalance both energies 
become equal and the energy of the Skyrmion
becomes lower than that of the pseudoSkyrmion at $\sigma >\sigma_c $. Therefore, 
at large imbalance e.g. $\sigma \simeq 1$, 
the Skyrmion excitation is
realized. Actually, the Skyrmion have been observed\cite{sawada} in the case that
all of electrons are involved only in one layer, i.e. $\sigma=1$.

We can distinguish these two types of Skyrmion excitations by observing
the dependence of the activation energy on parallel magnetic field. 
The energy of the pseudoSkyrmion decreases with the parallel magnetic field,
while the energy of the Skyrmion increases.
Thus, the distinction can be easily found.

First, we explain briefly Skyrmion excitations in  
single layer quantum Hall states by using the theory\cite{iwa1} 
of bosonized electrons, which is described by 
Chern-Simons gauge theory. Hamiltonian of bosonized electrons 
with spin is given by\cite{iwa,iwa1}

\begin{equation}
\label{h}
H=\int dx^2 \{\sum_{j=u,d}\frac{1}{2m_e}|(\partial_z -ia_z+ieA_z)\Psi_j|^2
+\frac{1}{2}g\mu B(|\Psi_d|^2-|\Psi_u|^2)\,\}
+\frac{\omega N}{2}
\end{equation}
with cyclotron frequency $\omega =eB/m_e$ and total number of
electrons $N$; we use the notation,
$\partial_z=\partial_1-i\partial_2$ and $a_z=a_1-ia_2 
\,\,(A_z=A_1-iA_2)$,
and use the unit of $\hbar=1$ and
light velocity $=1$. 
$\Psi_j$ represents field of bosonized electrons with spin
$j$ ( `$u$` and `$d$` denote spins parallel and anti-parallel to $\vec{B}$, 
respectively ).
$\vec{A}$ is a gauge potential 
of the magnetic field $\vec{B}$. 
$\vec{a}$ is a Chern-Simons gauge field satisfying a 
constraint,
$\partial_1 a_2-\partial_2 a_1 = 2\alpha (|\Psi_u|^2+|\Psi_d|^2)$.

The constraint equation implies 
that the bosonized field represents fermion
with the choice of $\alpha=\mbox{odd integer}\times \pi$: 
Attaching fictitious Chern-Simons flux of magnitude e.g. $2\pi$ to a boson makes the boson 
become a fermion.

We find from eq(\ref{h})
that the states $|L>$ satisfying the projection equation,
$(\partial_z -ia_z+ieA_z )\Psi_j|L>=0$ for all $j$,
are the states with the energy $\omega/2$ in the lowest Landau level. 
Hereafter, we analyze quantum Hall states 
in semiclassical 
approximation. It means that we solve classically
equations corresponding to 
these equations, that is, 
$(\partial_z -ia_z+ieA_z )\Psi_{u,d}(x)=0$ and
$\partial_1 a_2-\partial_2 a_1 = 2\alpha (|\Psi_u|^2+|\Psi_d|^2)$.

Solutions of the equations are 
given by, 
$\Psi_u=w_u(z)\exp(-a(x))$ and
$\Psi_d=w_d(z)\exp(-a(x))$ with $z=x_1-ix_2$,
where $a(x)$ is defined such as $a_i-eA_i=\epsilon_{ij}\partial_j
a(x)$ ( $\epsilon_{12}=1$ and $\epsilon_{ij}=-\epsilon_{ji}$ ) and 
satisfies the equation,
$-\partial^2a(x)=2\pi\{(|w_u|^2+|w_d|^2)\exp(-2a(x))-\rho\}$,
where $w_u$ and $w_d$ are arbitrary functions of $z$.
We have assumed $\alpha=\pi$ since we discuss only the states at
$\nu=1$ in this paper.
 
A solution of ground state 
describing uniform  
density of electrons $\rho$ is given by
$|w_u|=\sqrt{\rho}$ and $|w_d|=a(x)=0$.
It is obvious that the solutions, $|w_u|=\sqrt{\rho_u}$,
$|w_d|=\sqrt{\rho_d}$ 
and $a(x)=0$ with arbitrary $\rho_u$ and $\rho_d$,
are also possible ground states     
as far as $\rho_u+\rho_d=\rho$
when Zeeman energy vanishes.
It has, however, been argued that Coulomb interaction
and Fermi statistics of electrons ( 
namely, exchange interactions ) force the ground state
to be ferromagnetic ( only the state of electrons with parallel spin
component, $|w_u|=\sqrt{\rho}$ and $w_d=0$, is the true ground state
even without Zeeman energy).

Note that there is spin SU($2$) symmetry among the fields
in the absence of Zeeman energy.
Thus, this ferromagnetic state breaks spontaneously 
the symmetry, which gives rise to 
the phase coherence between up and 
down spin states, and solitons of Skyrmion.

A solution of such a Skyrmion is obtained by
choosing the functions $w_{u,d}$ such that
$w_u=\sqrt{\rho}\,z$ and $w_d=\sqrt{\rho}\,c$. 
Then, $a(x)$ satisfies 
$\label{a1}
-\partial^2 a(x)=2\pi\rho \{(r^2+c^2)\exp (-2a(x))-1\}$,
with $r^2=|z|^2$.
It turns out numerically that the parameter, $c$, represents
a typical scale of the spatial extention of the Skyrmion. 

The solution of the Skyrmion must satisfy a boundary 
condition, $|\Psi_u|\to \sqrt{\rho}$ as $r$ goes infinity.
This implies $\exp(-a(x))\to1/r$ as $r\to \infty$; we have found numerically
that the function, $\exp(-a(x))$, is regular at $r=0$
and a smooth decreasing function 
approximately given by $\sqrt{1/(r^2+c^2)}$ for $c \gg l$.
The electric charge of the Skyrmion,
is determined 
by this boundary condition such that
$\int dx^2 e\, ((|\Psi_u|^2-\rho)+|\Psi_d|^2)=-e >0$.

Zeeman energy of the Skyrmion is given by 
$-(1/2)\,g\,\mu \,B \int dx^2
  \{(|\Psi_u|^2-\rho)-|\Psi_d|^2\}=(1/2)\,g\mu B  
+g\mu B\rho\,c^2 \int dx^2\,\exp(-2a(x))$.
The Zeeman energy is proportional to the volume, $c^2$, of the Skyrmion and 
is logarithmically divergent since 
$\exp(-a(x))\to1/r$ as $r\to \infty$. But the integral
is finite when we take into account 
finiteness of actual coherent length, $\eta$,
of the boson field $\Psi_{u,d}$ \cite{ezawa}.
The length is qualitatively given by the inverse of Zeeman energy,
$(g\mu B)^{-1}$ since the presence of the
nonvanishing Zeeman energy breaks the SU($2$) symmetry explicitly
and makes the coherent length finite.

The length scale $c$ of the Skyrmion is determined by minimizing
the total energy of the Skyrmion, i.e.
$h_1e^2/\epsilon c\,+ g\mu B \, (c/l)^2 I(c)/2\pi$ where $I(c)=\int dx^2\,\exp(-2a(x))$ 
and $h_1\simeq 3\pi^2/64$. 
The Coulomb energy   
is approximately correct for $c$ larger than 
$l=\sqrt{1/eB}$.
The integral, $I(c)$, 
is almost constant in $c>l$; it behaves 
such as $I(c)\simeq 2\pi\log(\eta/c)$
for $c>l$. Thus, minimizing the energy
in $c$, we find the length scale of the Skyrmion,
$c_0\sim l\,\{(e^2/\epsilon l) /(g\mu B)\}^{1/3}>l$ 
and the energy of the Skyrmion,
$\sim (e^2/\epsilon l)\{(g\mu B)/(e^2/\epsilon l)\}^{1/3} > e^2/\epsilon l$.

We should mention that the Skyrmion has an additional energy
associated with the exchange interaction which gives rise to
the ferromagnetism. The energy is given by
$\int dx^2(\rho_s/2)\,\partial_k \vec{m}\partial_k \vec{m}=4\pi\rho_s$, 
where $\vec{m}$ denotes unit vector of the spin of
the Skyrmion, $\vec{m}\propto \Psi^{\dagger}\vec{\sigma}\Psi$
with $\Psi^{\dagger}=( \Psi_u^{\dagger}, \Psi_d^{\dagger})$.
$\rho_s=(1/16\sqrt{2\pi})e^2/\epsilon l$ is the spin stiffness
and $\vec{\sigma}$ are Pauli metrices.

We now proceed to discuss Skyrmion excitations in 
bilayer quantum
Hall states at $\nu=2\pi\rho/eB=1$;
$\rho=\rho_f+\rho_{\,b}$. 
In the bilayer system we
take into account indices distinguishing electrons 
located in different layers along with the spin indices (u, d).
We denote the field $\Psi_{fu}$ representing 
electrons with up spin located in the front-layer, $\Psi_{b\,u}$
representing electrons located in the back-layer, e.t.c.. 

All of these fields satisfy the 
following equations guaranteeing electrons lying
in the lowest Landau level, that is,
$(\partial_z -ia_z+ieA_z )\Psi(x)_j=0$ with
$\partial_1 a_2-\partial_2 a_1 = 2\pi (|\Psi_{fu}|^2+|\Psi_{fd}|^2+
|\Psi_{b\,u}|^2+|\Psi_{b\,d}|^2)$, 
where $j$ denotes one of four states of electrons. 
We should stress that there exists a local gauge symmetry, 
$\Psi_j\to \Psi_j\exp(i\Lambda )$
and $a_k \to a_k+\partial_k \Lambda$ in this system.

Solutions of the equations are given such that
$\Psi_j=w_j(z)\exp(-a(x))$
with arbitrary functions $w_j(z)$, 
where 
$a(x)$ is a solution of the equation, 
$-\partial^2 a(x)=2\pi\{(|w_{fu}|^2+|w_{fd}|^2+
|w_{b\,u}|^2+|w_{b\,d}|^2)
\exp(-2a(x))-\rho\}$. 

Ground states, as in the previous ferromagnetic case,
are given such that  
$\Psi_{fu}=\sqrt{\rho_f}\exp(i\theta_f)$ and 
$\Psi_{b\,u}=\sqrt{\rho_{\,b}}\exp(i\theta_{\,b})$ with any $\rho_l$
as far as $\rho=\rho_f+\rho_b$
( $ \Psi_{l\,d}=0$ for $l=f,b$
), where
$\theta_l$ are 
arbitral real parameters.      
This implies that quantum Hall states may
exist for any $\sigma$.
Actually, this result has been confirmed 
in the previous experiment\cite{sawada1} 
and is one of the outstanding 
properties of the quantum Hall state 
with the interlayer phase coherence. 
We mention that the difference of two phases $\theta_f-\theta_{\,b}$ in the state
is a physically relevant variable \cite{iwa} just like one in Josephson
junction, while the phase, $\theta_f+\theta_{\,b}$, 
is not physically relevant since the corresponding 
local gauge symmetry exists.

Now, we explain two types of Skyrmions \cite{ezawa};
Skyrmion and pseudoSkyrmion.
We address to only spherical 
form of solutions; namely $a(x)$ depends only on
$r$. First we show solutions of Skyrmions,
$w_{fu}=\sqrt{\rho_f}\,z, \quad w_{fd}=\sqrt{\rho_f}\,c,\quad 
w_{b\,u}=\sqrt{\rho_{\,b}}\,z,$ and $w_{b\,d}=\sqrt{\rho_{\,b}}\,c $.
Then, $a(x)$ satisfies 
$-\partial^2 a(x)=2\pi\rho \{(r^2+c^2)\exp(-2a(x))-1\}$,
with the boundary condition, $\exp(-a) \to 1/r$ as $r$ goes to infinity.

The Skyrmions have the same 
electric charge $-e$, Zeeman energy and 
exchange energy   
as Skyrmions in the single layer.
Contrary to the previous ones,
the charge is distributed over two layers,
$Q_k=-e\rho_k/\rho$ ( $-e=Q_f+Q_b$ ) with $k=f,b$. Furthermore,
the Skyrmions have tunneling energies whenever
$\Delta_{sas}$ is nonvanishing,
$E_t=-\frac{1}{2}\int dx^2\Delta_{sas}(\Psi_{fu}^{\dagger}\Psi_{b\,u}
+\Psi_{fd}^{\dagger}
\Psi_{b\,d}+c.c. 
-2 \sqrt{\rho_f\rho_b})=
\Delta_{sas}\frac{\sqrt{1-\sigma^2}}{2}$.

Therefore, the energy of the Skyrmion with $c > l$ is given by 

\begin{equation}
\label{Esk}
E_{sk}=4\pi\rho_s+\frac{h_1e^2}{\epsilon c}+
\frac{g\mu B}{2\pi} \,\frac{c^2}{l^2} I(c)+
\frac{\Delta_{sas}\sqrt{1-\sigma^2}}{2}+\frac{g\mu B}{2},
\end{equation}
where we have neglected the small charging
energy $\simeq 0.14 \sigma^2(e^2/\epsilon l)(l/c)(d/c)$, which
does not play important roles in the present discussion.

By minimizing $E_{sk}(c)$ in $c$, we find that the spatial extension, 
$c_0$, of the Skyrmion and $E_{sk}(c_0)$ are 
approximately given by

\begin{eqnarray}  
\label{es}
c_0 &\simeq&  l\,\{\frac{h_1e^2/2 \epsilon l}{g\mu B}\}^{1/3},\nonumber \\
E_{sk}(c_0)&\simeq&  4\pi\rho_s+
\frac{3h_1e^2}{2 \epsilon c_0}+\frac{\Delta_{sas}\sqrt{1-\sigma^2}}{2}+\frac{g\mu B}{2}, 
\end{eqnarray}
where we have assumed $\log(\eta/c)\simeq 1$ for simplicity\cite{ezawa}.
We note that
$E_{sk}(c_0)$ increases monotonously with Zeeman energy.

The other type of Skyrmions is pseudoSkyrmion
whose solutions are given such that
$w_{fu}=\sqrt{\rho_f}\,(z+c_f), \quad w_{fd}=0, \quad 
w_{b\,u}=\sqrt{\rho_b}\,(z+c_b)$ and $w_{b\,d}=0$,
where $c_i$ are arbitrary real parameters. 
We may choose the parameters $c_i$ without loss of generality such that
$\rho_f\,c_f+\rho_{\,b}\,c_{\,b}=0 \quad \mbox{and} \quad 
\rho_f c_f^2+\rho_{\,b} c_{\,b}^2=\rho\, c^2$.
Then, $a(x)$ satisfies 
$-\partial^2 a(x)=2\pi\rho \{(r^2+c^2)\exp(-2a(x))-1\}$.
We note that there are two centers in the solution ( $z=-c_f,\,-c_{\,b}$ )
and the phase $\theta_l$ of $\Psi_{lu}$ changes by $2\pi$ when we go around
the center at $z=-c_l$. Obviously, the parameter $c \propto |c_f-c_{\,b}|$ represents a length
scale of the phase coherence of the solution.

The pseudoSkyrmion has no electrons with
down spin component. Although it has no real spins,
the pseudoSkyrmion possesses pseudospin. Thus, it has
an exchange energy associated with the pseudospin,
$\int dx^2(\rho_{ss}/2)\,\partial_k \vec{m_s}\partial_k \vec{m_s}=4\pi\rho_{ss}$, 
where $\vec{m_s}$ denotes unit vector of 
the pseudospin;
$\vec{m_s}\propto \Psi_{ps}^{\dagger}\vec{\sigma}\Psi_{ps}$
with $\Psi_{ps}^{\dagger}=(\Psi_{fu}^{\dagger},\Psi_{b\,u}^{\dagger})$.
$\rho_{ss}$ is the pseudospin stiffness smaller than
$\rho_s$, in general; e.g.
$\rho_{ss}\simeq 0.14\rho_s(1-\sigma^2)$ for $d/l=1.5$\cite{mac}.

Contrary to the case of the Skyrmion, 
the pseudoSkyrmion possesses
a nontrivial charging energy because of
its nontrivial pseudospins $s_{3}(x)\propto m_{s,3}$ ,  

\begin{eqnarray}
&&\frac{e^2}{4\epsilon}\int dx^2dy^2(s_{3}(x)-s_{3})V_{-}(x-y)(s_{3}(y)-s_{3})\nonumber \\
&=&\frac{e^2}{4\epsilon}\int
dx^2dy^2\{\sigma^2\rho_r(x)\rho_r(y)+
(1-\sigma^2)\rho_t(x)\rho_t(y)\cos(\theta_x)\cos(\theta_y)\}V_{-}(x-y).
\end{eqnarray}
with $s_{3}(x)=(|w_{fu}|^2-|w_{b\,u}|^2)\exp(-2a(x))$,
$s_{3}=\rho_f-\rho_{\,b}$ and
$V_{-}=1/|x-y|\,-\,1/\sqrt{(x-y)^2+d^2}$,
where the integral with the coefficient, $\sigma^2$, is finite but the integral
with the coefficient, $1-\sigma^2$, is logarithmically divergent;
$\rho_{r}(x)=\rho\{(r^2-c^2)\exp(-2a(x))-1\}$ and 
$\rho_{t}(x)=2\rho\,c\,r\exp(-2a(x))$.
This divergent integral results from non spherical distribution
of the electric charge. Such charge distribution
would be screened. We expect that the screening 
effect makes the energy of this term much smaller 
than that of the first term proportional to $\sigma^2$ and  
the exchange ( Coulomb ) energy $\sim e^2/\epsilon l$.    
This assumption is crucial in our argument
but seems to hold in actual
samples because previous observations\cite{murphy,sawada1,sawada2} are 
consistent with our results based on the assumption.

The pseudoSkyrmion also has the tunneling energy,
$E_t=\Delta_{sas}\sqrt{1-\sigma^2}/2+
\Delta_{sas}\sqrt{1-\sigma^2}\,c^2 I(c)/2\pi l^2$.
Therefore, the total energy of the pseudoSkyrmion with $c > l$
is given by

\begin{equation}
E_{psk}=4\pi\rho_{ss}+\frac{h_1e^2}{\epsilon c} +\frac{h_2e^2\sigma^2}{\epsilon l}\,\frac{d}{l}\,
\,\frac{c^2}{l^2}  
+\frac{\Delta_{sas}\sqrt{1-\sigma^2}}{2\pi}\,\frac{c^2}{l^2}\,I(c)
+\frac{\Delta_{sas}\sqrt{1-\sigma^2}}{2} +\frac{g\mu B}{2},
\end{equation}
with $h_2\simeq 0.4$.
We find that  
the charging energy is much larger than that of the Skyrmion,
which makes us distinguish experimentally the Skyrmion
from the pseudoSkyrmion.

By minimizing $E_{psk}$ in the parameter $c$, we find that 
the length scale $c_0$ and the energy of the pseudoSkyrmion are 
approximately given by

\begin{eqnarray}
\label{eps}
c_0&\simeq& l\, \{\frac{h_1e^2/2 \epsilon l}{\Delta_{sas}\sqrt{1-\sigma^2}+
(h_2e^2\sigma^2/\epsilon l)(d/l)}\}^{1/3},\nonumber \\
E_{psk}(c_0)&\simeq& 4\pi\rho_{ss}+\frac{3 h_1e^2}{2 \epsilon \,c_0}
+\frac{\Delta_{sas}\sqrt{1-\sigma^2}}{2}+\frac{g\mu B}{2}.
\end{eqnarray}

Roughly speaking, $E_{psk}(c_0)$ 
decreases with the tunneling energy, 
$\Delta_{sas}$, for the small imbalance,  
$\sigma^2 \ll 1$, while
it remains a constant for $\sigma^2 \simeq 1$.

We comment that the terms, $\Delta_{sas}\sqrt{1-\sigma^2}/2+g\mu B/2$
in both $E_{sk}$ and $E_{psk}$ are irrelevant 
in the observation of activation energies.
A pair of Skyrmion and anti-Skyrmion is actually observed so that 
these terms cancel with corresponding ones of anti-Skyrmion, 
 $-(\Delta_{sas}\sqrt{1-\sigma^2}/2+g\mu B/2)$. Thus, we ignore
the terms in the subsequent discussions.   

Here we should mention that the tunneling energy 
of both types of Skyrmions decreases by increasing
parallel magnetic field, $B_{\|}$,
with keeping longitudinal component $B_{\bot}$ ( $\nu=2\pi/eB_{\bot}=1$ ),

\begin{eqnarray}
\label{Etsk}
E_t^{B_{\|}}(\mbox{Sk})&=&-\Delta_{sas}^Q\frac{\sqrt{1-\sigma^2}}{2}\,
\rho\int dx^2\{(r^2 + c^2)\exp(-2a(x))-1\}\cos(Qx_1) \\
E_t^{B_{\|}}(\mbox{pSk})&=&E_t^{B_{\|}}(\mbox{Sk}) 
         +2\Delta_{sas}^Q\frac{\sqrt{1-\sigma^2}}{2}\,
\rho\int dx^2 c^2\exp(-2a(x))\cos(Qx_1)
\label{Etpsk}
\end{eqnarray} 
with $Q=edB_{\|}$,
where `Sk` for Skyrmion (`pSk` for pseudoSkyrmion )
and these replace the tunneling energies in eq(\ref{es}) and
eq(\ref{eps}).
The tunneling strength,
$\Delta_{sas}$, is reduced to
$\Delta_{sas}^Q=\Delta_{sas}\exp(-(d/2l)^2\tan^2\Theta)$ owing to
suppression of tunneling 
by the parallel magnetic field; $\tan\Theta=B_{\|}/B_{\bot}$. 
We can see numerically that 
this tunneling energy becomes small with $Q$ and 
vanishingly small when $Q > c^{-1}$ owing to the factor,
$\cos(Qx_1)$. Note that $c$ is determined for the energies of
Skyrmions to be minimized. Especially 
in the case of the pseudoSkyrmion at $\sigma=0$, 
such $c(Q)$ increases with $Q$
since $c(Q)$ increases as $E_t^{B_{\|}}$ decreases
and $E_t^{B_{\|}}$ decreases with $Q$. But,
$c(Q)$ can not be larger than the coherent length 
$\eta\propto \Delta_{sas}^{-1}$ of the phase,
$\theta_f-\theta_{\,b}$.
Thus, there exist a critical point $Q_c \simeq c(Q_c)^{-1}=\eta^{-1} 
\propto \Delta_{sas}$, at which
the tunneling energy becomes vanishingly small,

\begin{equation} 
\label{t}
Q_c=edB_{\|}\propto \Delta_{sas} \quad \mbox{or} \quad \tan \theta_c
=h_0\,\frac{l}{d}\, \frac{\Delta_{sas}}{e^2/l}
\end{equation}  
with $h_0$ being a numerical constant. Hence, the 
activation energy of the pseudoSkyrmion decreases with $Q$ 
until $Q$ is equal to $Q_c$.
( The ground state has been argued to change from 
a commensurate state to an incommensurate state 
when $Q$ goes beyond the critical 
point $Q_c$. Thus, it is necessary to analyze 
the relevance of these solitons in the region of $Q>Q_c$ more carefully.)
This estimation in eq(\ref{t}) gives a rough agreement 
with previous measurements\cite{murphy}  
of the activation energies. On the other hand,
in the case of Skyrmion the scale
$c_0\propto (g\mu B)^{-1/3}$ does not depend
on $E_t^{B_{\|}}$. Thus,
the dependence of its energy on $B_{\|}$ is rather simple;
there is no critical behavior in $Q$.

Finally, we wish to discuss which type of Skyrmion is realized 
when we change the imbalance parameter.
We find that at $\sigma=0$
the energy of the pseudoSkyrmion 
is smaller than that of the Skyrmion
because $\rho_{ss}\ll \rho_{s}$, while 
at $\sigma=1$ the energy of the Skyrmion is smaller than that 
of the pseudoSkyrmion. 
This is because  
the charging energy of the pseudoSkyrmion vanishes at $\sigma=0$ 
but becomes maximal, i.e. $\sim (e^2/\epsilon l)d\,c_0^2/l^3$ 
at $\sigma=1$.
Thus there exists a critical point
$\sigma_c$, at which a transition between the Skyrmion and
the pseudoSkyrmion arises; e.g. $\sigma_c\simeq 0.4$,  
with a typical sample parameter, $\rho=10^{11}/\mbox{cm}^2$,
$d/l=1.5$ and
$\Delta_{sas}= 1\sim 5$K.
This agrees roughly with the experiment \cite{sawada2}.

The transition between the Skyrmion and the pseudoSkyrmion
can be seen by
the measurement of the dependence of the activation energy
on $B_{\|}$ or $\sigma$. 
Imposing the parallel magnetic
field with fixing $\nu=1$,
we make Zeeman energy increase, but 
the tunneling energy, $E_t^{B_{\|}}$, decrease. Therefore,
The energy 
of the Skyrmion increases monotonously with $B_{\|}$
such as $E_{sk}(c_0)\propto (g\mu B)^{1/3}$. 
On the other hand, 
the energy, $E_{psk}(c_0)\propto 1/c_0(Q)$, 
of the pseudoSkyrmion decreases with $B_{\|}$, but
remains a constant when $B_{\|}>Q_c/ed$;
$c_0(Q)$ can not be larger than $\eta \sim \Delta_{sas}^{-1}$.
Furthermore,
we find that the energy of the pseudoSkyrmion increases
with $\sigma$ for $\sigma < \sigma_c$, while that of the Skyrmion
remains a constant for $\sigma > \sigma_c$; more precisely, owing to
the presence of 
the charging energy, $\simeq 0.14 \sigma^2(e^2/\epsilon l)(l/c)(d/c)$, 
of the Skyrmion, the total energy of the Skyrmion increases but much 
more slowly with $\sigma$ than that of
the pseudoSkyrmion increases. Thus, the distinction of these two types of Skyrmions   
can be easily found.
The dependence of the activation energies on $\sigma$ and $B_{\|}$  
has been partially measured \cite{murphy,sawada1,sawada2}
and is consistent with our present discussions. 
 
Probably, Skyrmions and pseudoSkrmions are  
rellevant excitations
in the bilayer quantum Hall states with the interlayer phase
coherence
at $\nu=1$. In order to confirm our results, 
an experiment suggested in this paper
is in progress.


We thank F.Z. Ezawa and A. Sawada for fruitful discussions 
and members of theory group in KEK for their
hospitality.





\end{document}